\documentclass[preprint]{aastex}

\slugcomment{Submitted to PASP on 2000 Aug 30; accepted on 2000 Oct 06}

\shorttitle{Distance to V716 Mon}
\shortauthors{Hoard et al.}

\begin{document}

\title{Distance to the RR Lyrae Star V716 Monocerotis\altaffilmark{1}}

\author{D. W. Hoard\altaffilmark{2}}
\affil{Cerro Tololo Inter-American Observatory, Casilla 603, La Serena, Chile}
\email{dhoard@noao.edu}
\authoraddr{CTIO, P.O. Box 26732, Tucson AZ 85726}

\author{Andrew C. Layden}
\affil{Physics \& Astronomy Dept., Bowling Green State Univ., Bowling
Green, OH, 43403}
\email{layden@baade.bgsu.edu}
\and
\author{Jeremy Buss\altaffilmark{3}, Ricardo Demarco\altaffilmark{4}, Jenny Greene\altaffilmark{5}, Jessica Kim-Quijano\altaffilmark{6}, Alicia M. Soderberg\altaffilmark{7}}
\affil{Cerro Tololo Inter-American Observatory, Casilla 603, La
Serena, Chile}

\altaffiltext{1}{This research was conducted as part of the 1999 Research Experiences for Undergraduates (REU) and Pr\'{a}cticas de Investigaci\'{o}n en Astronom\'{i}a (PIA) Programs at Cerro Tololo Inter-American Observatory (CTIO).}
\altaffiltext{2}{Director of the CTIO REU and PIA Programs.}
\altaffiltext{3}{REU student from University of Wisconsin-Oshkosh.}
\altaffiltext{4}{PIA student from Pontificia Universidad Cat\'{o}lica de Chile, Santiago.}
\altaffiltext{5}{REU student from Yale University.}
\altaffiltext{6}{REU student from Towson University/Space Telescope Science Institute.}
\altaffiltext{7}{REU student from Bates College.}

\begin{abstract}

We present high quality $BVRI$ CCD photometry of the variable star
V716 Monocerotis (= NSV 03775).  We confirm it to be an RR
Lyrae star of variability type ab (i.e.\ a fundamental mode pulsator), and determine its metallicity ([Fe/H] $= -1.33 \pm 0.25$), luminosity ($M_V = 0.80 \pm 0.06$), and
foreground reddening ($E(B-V) = 0.05$--$0.17$) from the Fourier components of its light curve.  These parameters indicate a distance of $4.1 \pm 0.3$ kpc, placing V716 Mon near the plane of the
Galaxy well outside the solar circle.

\end{abstract}


\section{Introduction}

The classification of variable stars can often depend sensitively on the available observational data.  For example, the periodic rapid brightness increase displayed by RR Lyrae stars can easily be confused with typical variability of cataclysmic variables (such as the start of a dwarf nova outburst) if only limited observations are available.  A number of alleged cataclysmic variables (CVs) have been reclassified as RR Lyrae stars following more extensive observation \cite[e.g.\ CG Muscae,][]{LW97}.  As RR Lyrae variables are particularly useful objects for exploring Galactic structure and dynamics, owing to the availability of relationships between their observational properties and their distances, it is desirable to continue to reclaim misclassified RR Lyrae stars from the ranks of the CVs.

V716 Monocerotis (= NSV 03775) was first identified as a variable star by \citet{hoffmeister49}.  It was categorized as a suspected CV or RR Lyrae variable by \citet{KK82}.  \citet{KO95} ruled out the possible CV nature of V716 Mon by obtaining a spectrum (indicative of an A-type star) and several CCD and photoelectric data sets (showing an almost complete cycle of the characteristic light curve of an RR Lyrae star).  \citet*{haefner96} independently concluded that V716 Mon was an RR Lyrae star based on a steady decrease in its relative brightness of $\approx0.44$ mag during a 4.75 hr CCD light curve.  The CV catalog and atlas of \citet*{DWS97} now lists V716 Mon as a ``non-CV,'' thus providing the final word in the tale of its long-running identity crisis. 

We observed V716 Mon during the telescope orientation of the 1999 Cerro Tololo Inter-American Observatory (CTIO) summer student programs.  We chose this object for two main reasons: (1) no calibrated photometry currently exists for it in the literature, and (2) it is located in a field that is not too crowded to perform reliable photometry at the plate scale of the CTIO Curtis Schmidt telescope ($2\farcs32$ pixel$^{-1}$ -- see \S\ref{s-obs}).
We present here the first calibrated $BVRI$ photometry of V716 Mon.  We use our multi-color light curves to calculate several fundamental parameters of this RR Lyrae star, including its distance and position within the Galaxy.

\section{Observation}
\label{s-obs}

We observed V716 Mon for 3 nights on 1999 January 27--29 UT using a $2048\times2048$ SITe CCD on the Curtis Schmidt telescope at CTIO\footnote{See \url{http://www.astro.lsa.umich.edu/obs/schmidt/}}.  We utilized $BVRI$ filters with exposure times of 40 s, 20 s, 10 s, and 10 s, respectively.  The filters were cycled in the sequence $BVRIIRVB$ in order to minimize dead time due to filter wheel movement and obtain near-simultaneous multi-color light curves.  We defined a region-of-interest $1024\times1024$ pixels in size and used a single amplifier for the CCD read-out.  Including filter changes, this resulted in a typical cycle time of $\approx30$ s between exposures.  Table \ref{t-log} lists the number of images and time coverage obtained on each night.

The images were reduced in the normal fashion using standard IRAF\footnote{The Image Reduction and Analysis Facility software, which is distributed by the National Optical Astronomy Observatories.} tasks.  Instrumental magnitudes were measured using the IRAF task {\em phot}.  We obtained observations of photometric standard stars \citep{Land92} on Jan 27 \& 29.  The standard star observations from Jan 27 (the better night) were checked against those from Jan 29, and the former were used to calibrate the Jan 27 data for V716 Mon and several field stars.  These calibrated field stars were then used as secondary standards to calibrate the V716 Mon data on Jan 28 \& 29.  This gave photometric uncertainties on Jan 27 of $\sigma_{B} = 0.02$ mag and $\sigma_{V} = \sigma_{R} = \sigma_{I} = 0.01$ mag.  The uncertainties on Jan 28 \& 29 were $\sigma_{B} = 0.03$ mag and $\sigma_{V} = \sigma_{R} = \sigma_{I} = 0.02$ mag.

\section{Analysis}

\begin{figure}[tb]
\plotone{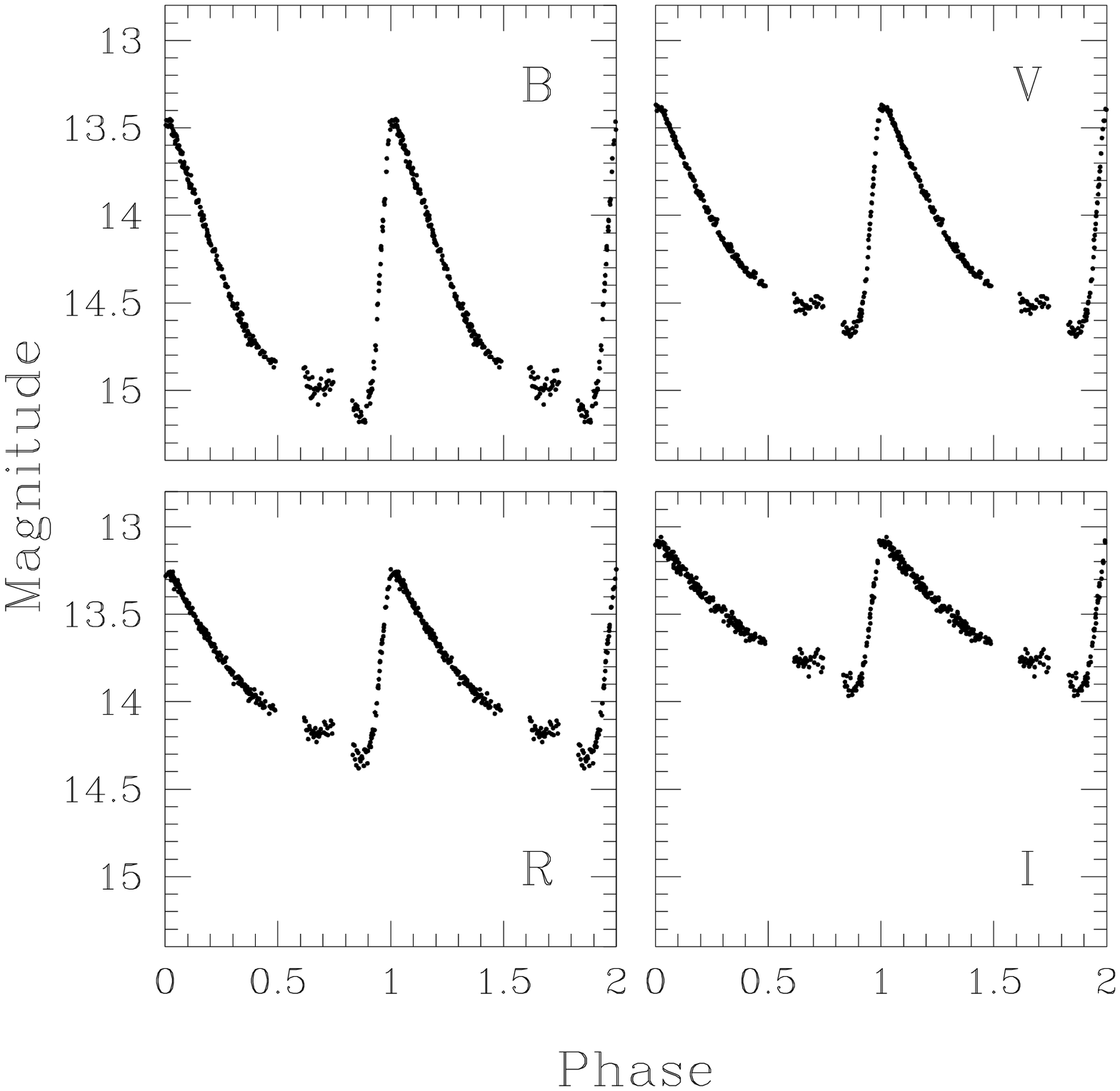}
\caption{Phased light curves ($P = 0.565$ d) of V716 Mon in the $B$, $V$, $R$, and $I$
bandpasses.  All four light curves are shown on the same magnitude scale, and the data are plotted twice for clarity.  \label{f-4lcs}}
\end{figure}

\citet{KO95} found a period of 0.565 d for V716 Mon from their CCD and photoelectric photometry data.  We applied the phase dispersion minimization technique \citep{Stel78} to our data, and confirm this period.
Figure \ref{f-4lcs} shows the photometry data folded on $P = 0.565$ d.
Clearly, V716 Mon is an ab-type RR Lyrae variable.  Table \ref{t-lcpars}
presents light curve parameters derived from the $BVRI$ light curves
shown in Figure \ref{f-4lcs}.  These parameters include the intensity-mean magnitude ($\langle M \rangle$), the magnitudes at maximum light ($M_{max}$) and at
minimum light ($M_{min}$), and the phase difference between minimum
and maximum light ($\Delta\phi_{rise}$).  The magnitudes are precise
to within a few hundredths of a magnitude.

%

Using our data, we can derive a number of physical properties of
V716 Mon, and thereby determine both its location within the Galaxy and to
which stellar population it belongs.  Fourier decomposition of a
star's light curve is a useful tool in this regard.  It yields an
amplitude and phase shift for each sinusoid (order) employed in the
fit.  For RR Lyrae stars, several empirical relations have been
developed between these Fourier components and the stars' physical
properties \cite[metallicity, luminosity, color, etc.;][]{jk96,kj96,kj97}.

We used code kindly provided by Jurcsik \& Kov\'{a}cs to perform
Fourier decompositions of the V716 Mon light curves in a manner
consistent with that used to develop the empirical relations.
\citet{jk96} (hereafter referred to as JK96) used fits
involving 15 Fourier orders.  Using such high orders to fit our data
yielded significant excursions (``wiggles'') within the phase gaps at
$\phi = 0.55$ and 0.78 that do not appear in the light curves of other
observed RR Lyrae stars.  The excursions must be spurious artifacts.  We
therefore explored using between four and fifteen Fourier orders in
our fits to the $V$-band light curve.

\begin{figure}[tb]
\plotone{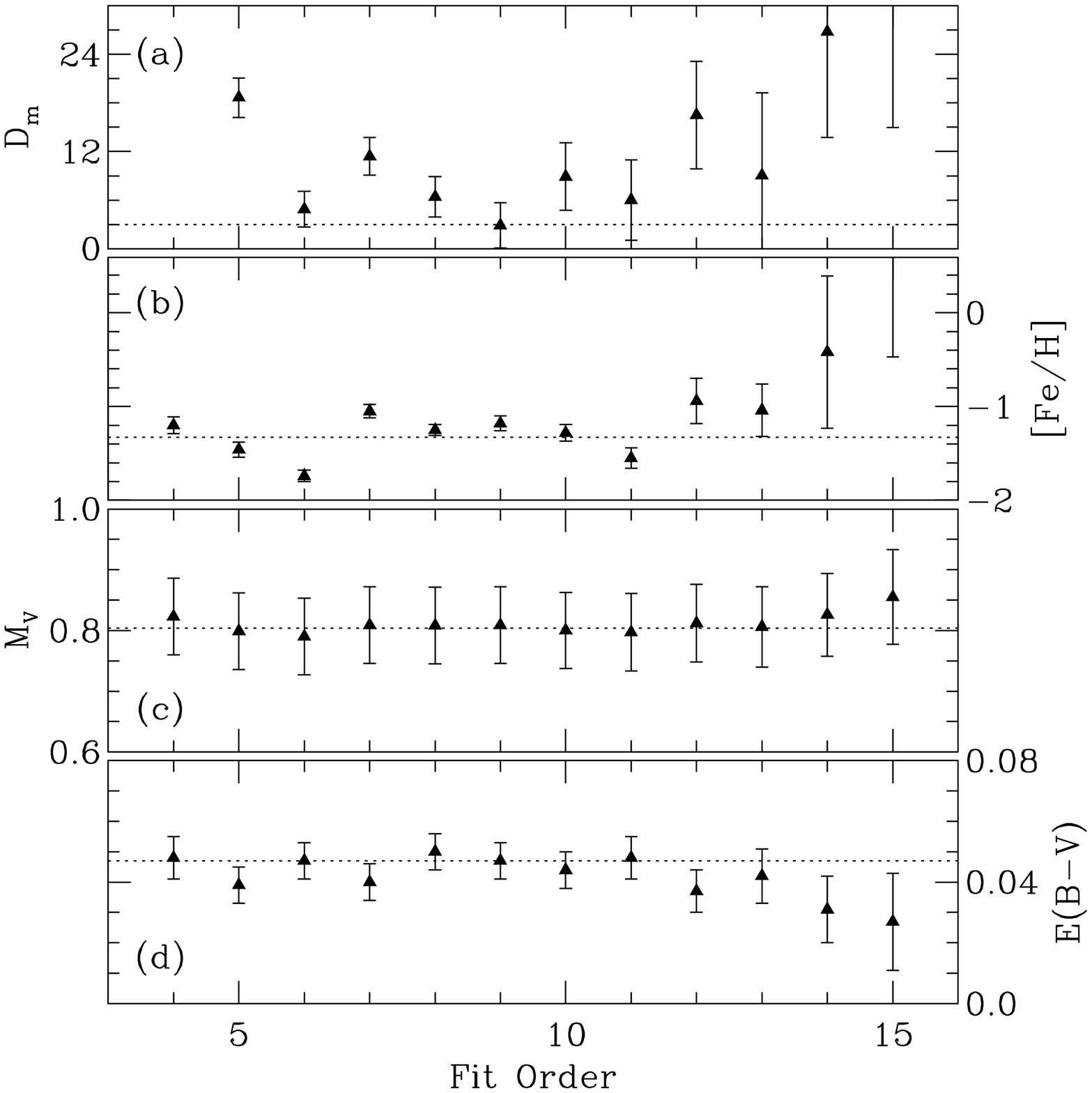}
\caption{Results of the Fourier fits to the V716 Mon light curves.
(a) The compatibility condition ($D_m$) of JK96 is plotted as a
function of the number of Fourier components used in the 
fit (i.e.\ the fit order).
Fits with $D_m < 3$ (dotted line) are deemed to be compatible with the
data. The estimated metallicity (b), absolute magnitude (c), and 
interstellar reddening (d) are plotted against fit order.
The results are best determined using fit orders between 6 and 11.
The dotted lines in (b)--(d) are the adopted, weighted-mean values
based on these fits.  \label{f-two}}
\end{figure}

JK96 provided a compatibility test to determine whether an individual
fit was reliable based on the value of a parameter, $D_m$.  This is the 
maximum of the deviation parameters that are defined as the differences 
between observed and calculated Fourier parameters divided by the 
standard deviation of the Fourier parameter fits (see Section 4 of JK96).  
If $D_m < 3$, the fit was deemed compatible with
the data.  Figure~\ref{f-two}a shows our $D_m$ values as a function of the fit
order, $O_F$, with the dotted line marking $D_m = 3$.  Only one fit
($O_F = 9$) formally 
satisfies the compatibility test, but fits with $6 < O_F <
11$ come close to meeting the test.  When plotted over the data, these
fits (in particular, $O_F = 9$) do a reasonable job of representing
the observations.  As we will show, the results based on these fits
are similar, providing confidence that we have obtained Fourier
deconvolutions that are meaningful in the context of determining the
physical properties of V716 Mon.

JK96 derived an empirical relation between metallicity ($[Fe/H]$),
period, and the phase shift $\phi$$_{3}$$_{1}$ (their Eqn.\ 3).
Employing this relation, we obtained the $[Fe/H]$ values shown in
Figure~\ref{f-two}b from the Fourier fits with 4--15 orders.  Notice that the
results for $6 < O_F < 11$ are clustered together and have small
internal errors.  We therefore adopt as our best $[Fe/H]$ estimate the
weighted mean of the individual $[Fe/H]$ values derived from fits with
$6 < O_F < 11$, where the weights are $(1/D_m)^2$.  Since the results
for individual fit orders are not independent, we estimate the error
in $[Fe/H]$ from the range of individual values ($-1.05$ to $-1.55$).
We thus obtain $[Fe/H] = -1.33 \pm 0.25$ for V716 Mon.

\citet{kj96} derived a similar empirical relationship
between absolute magnitude, period, $\phi$$_3$$_1$, and the Fourier
amplitude $A_1$.  
\citet{kj97} (hereafter referred to as KJ97) improved this relation and
extended it to include a reddening estimate.  Figure~\ref{f-two}c shows the
absolute magnitude 
results from the fits with 4--15 orders.  The results are highly
consistent among the fits that used $6 < O_F < 11$, so we again adopt
as our best estimate the weighted mean of the individual $M_V$ values
derived from these fits (using the inverse-square of $D_m$ for the
weights).  The range of individual $M_V$ values (0.790 to 0.809 mag)
is much smaller than the internal error in an individual estimate
(0.063 mag), so we adopt the latter as our error estimate on $M_V$.
Thus we find $M_V = 0.80 \pm 0.06$ mag for V716 Mon.  The KJ97
absolute magnitude relation was calibrated using Baade-Wesselink
luminosities.  Other luminosity calibrations could result in a
systematic decrease in $M_V$ by as much as 0.3 mag \citep{chaboyer99}.

Again using the relations in KJ97, we determined the reddening from
fits with $4 < O_F < 15$ (see Figure~\ref{f-two}d).  For $6 < O_F < 11$
the results are nearly independent of the fit order, so we adopt as
our best reddening estimate the weighted mean of these values.  For
the error we use the quadrature sum of the external error (estimated
from the range of values, 0.040 to 0.050 mag) and the internal error
(estimated from the errors in the Fourier coefficients, 0.005 mag).
Thus $E(B-V) = 0.05 \pm 0.01$ mag for V716 Mon.

We can compare this reddening estimate with several independent
methods.  \citet{blanco92} developed a semi-empirical relation between
$E(B-V)$ and the period, metallicity, and minimum-light $(B-V)$ colors
for ab-type RR Lyrae stars.  \citet{muskkk95} explored a similar relation
involving $(V-I)$ colors.  Over the phase interval 0.5 to 0.8, the
average colors of V716 Mon are $(B-V) = 0.46 \pm 0.01$ mag and $(V-I)
= 0.74 \pm 0.01$ mag.  The \citet{blanco92} and \citet{muskkk95}
methods yield $E(B-V) = 0.07 \pm 0.01$ and $0.12 \pm 0.02$ mag,
respectively.  The dust map of \citet{schlegel98} yields a larger
value of $E(B-V) = 0.17 \pm 0.02$ mag.  Despite its proximity to the
plane ($b = +8.41$ deg), the reddening toward V716 Mon is surprisingly
low.

We compute the distance to the star using the absolute magnitude and
reddening values determined above, along with the average apparent
magnitude derived from the Fourier fits ($V = 14.20 \pm 0.02$ mag).
Using the low reddening value of 0.05 yields a distance of $4.46 \pm
0.15$ kpc, while using the reddening from the \citet{schlegel98} maps
gives $3.76 \pm 0.15$ kpc.  Given this uncertainty, we adopt $d = 4.1
\pm 0.3$ kpc.  Systematic changes in the assumed absolute magnitude
zero-point scale this distance.
The galactic coordinates of the star are $(l, b) = (229.61, +8.41)$ deg.
At the adopted distance, V716 Mon lies 0.6 kpc above the
Galactic plane and 11.1 kpc from the Galactic center (assuming $R_0 =
8$ kpc).  Given its metallicity and location, it could be a member
either of the Galaxy's halo population, or of the metal-weak tail of
the thick disk population \citep{layden95}.  Radial velocity
measurements and spectroscopic confirmation of the metallicity would
likely remove this ambiguity.

\section{Conclusions}

We have obtained extensive $BVRI$ photometry of the variable star
V716 Mon.  Its period and multi-color light curves indicate that it is an RR Lyrae
variable of Bailey type ab (i.e.\ a fundamental mode pulsator).
Using the empirical relations between light curve Fourier components
and physical properties (JK96, KJ97), we estimate the star's
metallicity to be [Fe/H] $= -1.33 \pm 0.25$ dex, its absolute magnitude
to be $M_V = 0.80 \pm 0.06$ mag (internal error only), and its
reddening to be in the range $E(B-V) = 0.05$--$0.17$ mag.  The resulting
distance is $4.1 \pm 0.3$ kpc, placing V716 Mon near the plane of the
Galaxy well outside the solar circle.

\acknowledgements

We thank the anonymous referee, whose comments prompted an improvement in the analysis of these data.  
The CTIO REU Program is funded by the National Science Foundation (NSF).  The CTIO PIA Program is funded by CTIO.  CTIO is operated by AURA, Inc.\ under cooperative agreement with the NSF.  This research made use of NASA's Astrophysics Data System Abstract Service and the SIMBAD database operated by CDS, Strasbourg, France.


\begin{deluxetable}{lllcccc} 
\tablewidth{0pt}
\tablecaption{Log of Observations \label{t-log}} 
\tablehead{
\colhead{ }    &  
\multicolumn{2}{c}{Time Coverage} &
\multicolumn{4}{c}{Number of Images} \\
\colhead{UT Date} &
\colhead{UT} &
\colhead{HJD-2451000} &
\colhead{$B$}    &  
\colhead{$V$}    &  
\colhead{$R$}    &  
\colhead{$I$}    
}
\startdata
1999 Jan 27 & 02:44--08:05\tablenotemark{a} & 205.6197--205.8791 & 51 & 52 & 52 & 52 \\
1999 Jan 28 & 02:28--09:00 & 206.6086--206.8779 & 101 & 101 & 101 & 101 \\
1999 Jan 29 & 01:38--08:30\tablenotemark{b} & 207.5741--207.8572 & 103 & 102 & 102 & 101
\enddata 
\tablenotetext{a}{Less 0.5 hr used for standard stars.}
\tablenotetext{b}{Less 0.75 hr used for standard stars.}
\end{deluxetable} 
\begin{deluxetable}{ccccc} 
\tablewidth{0pt}
\tablecaption{Light Curve Parameters \label{t-lcpars}} 
\tablehead{
\colhead{Parameter}    &  
\colhead{$B$}    &  
\colhead{$V$}    &  
\colhead{$R$}    &  
\colhead{$I$}  
}
\startdata
$\langle M \rangle$   & 14.47  & 14.15  & 13.86  & 13.52  \\
$M_{max}$             & 13.47  & 13.38  & 13.26  & 13.08  \\
$M_{min}$             & 15.16  & 14.67  & 14.34  & 13.94  \\
$\Delta \phi _{rise}$ & 0.14   & 0.15   & 0.14   & 0.14   
\enddata 
\end{deluxetable} 


\begin{thebibliography}{}

\bibitem[Blanco(1992)]{blanco92}Blanco, V. M. 1992, \aj, 104, 734

\bibitem[Chaboyer(1999)]{chaboyer99}Chaboyer, B. 1999, in Post-{\em Hipparcos} Cosmic Candles, eds. A. Heck, F. Caputo (Dordrecht: Kluwer Academic Publ.), 111

\bibitem[Downes et al.(1997)Downes, Webbink, \& Shara]{DWS97} Downes, R., Webbink, R. F., \& Shara, M. M. 1997, \pasp, 109, 345 

\bibitem[Haefner et al.(1996)Haefner, Fiedler, \& Rau]{haefner96}Haefner, R., Fiedler, A., \& Rau, S. 1996, IBVS, 4366

\bibitem[Hoffmeister(1949)]{hoffmeister49}Hoffmeister, C. 1949, Astr. Abh., 12, No. 1

\bibitem[Jurcsik \& Kov\'{a}cs(1996)]{jk96}Jurcsik, J., Kov\'{a}cs, \& G. 1996, \aap, 312, 111 (JK96)

\bibitem[Koen \& O'Donoghue(1995)]{KO95}Koen, C., \& O'Donoghue, D. 1995, \apjs, 101, 347

\bibitem[Kov\'{a}cs \& Jurcsik(1996)]{kj96}Kov\'{a}cs, \& G., Jurcsik, J. 1996, \apjl, 466, L17 (KJ96)

\bibitem[Kov\'{a}cs \& Jurcsik(1997)]{kj97}Kov\'{a}cs, G., \& Jurcsik, J. 1997, \aap, 322, 218 (KJ97)

\bibitem[Kukarkin \& Kholopov(1982)]{KK82}Kukarkin, B. V., \& Kholopov, P. N. 1982, New Catalogue of Variable Stars (Moscow: Publication Office Nauka)

\bibitem[Landolt(1992)]{Land92}Landolt, A. U. 1992, \aj, 104, 340

\bibitem[Layden(1995)]{layden95}Layden, A. C. 1995, \aj, 110, 2288

\bibitem[Layden \& Wachter(1997)]{LW97}Layden, A. C., \& Wachter, S. 1997, \pasp, 109, 977

\bibitem[Mateo et al.(1995)]{muskkk95}Mateo, M., Udalski, A., Szymanski, M., Kaluzny, J., Kubiak, M., \& Krzeminski, W. 1995, \aj, 109, 588

\bibitem[Schlegel et al.(1998)Schlegel, Finkbeiner, \& Davis]{schlegel98}Schlegel, D. J., Finkbeiner, D. P., \& Davis, M. 1998, \apj, 500, 525

\bibitem[Simon \& Teays(1982)]{ST82}Simon, N. R., \& Teays, T. J. 1982, \apj, 261, 586

\bibitem[Stellingwerf(1978)]{Stel78}Stellingwerf, R. F. 1978, \apj, 224, 953

\end{thebibliography}
\end{document}